\documentclass[aps,showpacs,nobibnotes,showkeys,superscriptaddress,
eqsecnum,amssymb,epsfigs,amsfonts,nofootinbib,amsmath,apsfonts]{revtex4}
\usepackage{graphicx}
\usepackage{dcolumn}
\usepackage{bm}
\begin{document}
\title{\bf\large Four Dimensional Superstring, Supergravity and Duality}
\author{B. B. Deo\footnote{E-mail:bdeo@iopb.res.in}
and L Maharana\footnote{E-Mail:lmaharan@iopb.res.in}}
\affiliation{Department of Physics, Utkal University, Bhubaneswar-751 004, India}
\begin{abstract}
Montonen and Olive's conjecture in 1977 that some theories 
possess  a duality symmetry that 
interchanges the electrically charged particle with that of negatively charged 
t' Hooft-Polyakov monopoles which relates strong coupling to weak coupling. 
This will be nearly  the same as Shapere et al, who showed
that the equation of motion of the coupled Einstein-Maxwell-axion dilaton
system, occurs in low energy in String theory, is invariant under 
electromagnetic duality transformation; that also interchanges the strong 
and weak coupling limit of the theory. These ideas have been persued in the 
$SO(6)\otimes SO(5)$ gauge groups as well as in the supersymmetric group of
a four dimensional String. We analyse and find that they give 
clearer evidence of string theory in two and four dimensional 
supergravity theory. However, the application to supergravity is beset with
the normal difficulties and suggestions have been made how to settle  them.
\end{abstract}
\date{\today}
\maketitle
\section{Introduction}\label{int}
The classical relativistic string theory, first proposed by Nambu~
\cite{Nambu70} and Goto~\cite{Goto71}, turned out to be 
valid only in 26  dimensions and was raised to the quantum level by
Goddarda, Goldstone, Rebbi, Thorn\cite{Goddard}
and Mandelstam~\cite{Mandelstam75}. However, Scherk and Schwarz~\cite{Sherk}
pointed out that the string theory, in 26 dimensions, can explain the physical phenomena 
including gravity, but this did not make much  progress. It was shown by Deo and 
collaborators~\cite{Deo03,Deo04,Deo05,Deo06} that a four dimensional 
superstring can easily be constructed and is anomally free, modular 
invariant, free from ghosts and  unitary. The results are those obtained
with the group SO(3,1) and $SO(6)\otimes SO(5)$. These have been enumerated 
in detail in the papers~\cite{Deo03,Deo04,Deo05,Deo06}. However, we shall give some of 
the features in the next Section-\ref{superstring}.

Like all string theories, the theory had the defect of on being unable to explain the
other important features like gauging, duality and supersymmetry, 
relevant to supergravity theory. Our main purpose is to consider these problems 
and try to find the solutions.
In the Section-\ref{invariant}, we shall discuss  the main features of
duality as it enters in the domain of superstring theories.
Here we use the element of supergravity theory and apply it to the 
superstring theory. Montonen\cite{Montonen77} in 1977 has conjectured that the
spontaneously broken gauge symmetry possesses a duality symmetry  that 
interchanges electrically charged particles with negative charged 
t'~Hooft-Polyakov monopoles. Such symmetry  relates strong to weak coupling,
because we set the coupling constant to its inverse. The best instance is 
found in relating the Montonen-Olive duality to globally supersymmetric 
Yang-Mills-Higgs system. More recently,  Shapere et al~\cite{Shapere91} 
observed and showed that the equation of motion of coupled  Einstein-Maxwell's action,
which is the effective action in string theory in low energy, is invariant under
electromagnetic duality transformation and  also interchanges the strong and
weak coupling limit of the string ttheory.

It is very  important and useful to construct the dual description of 
theories, containing
antisymmetric tensor fields by opening up a new potential, whose curl gives the dual
field strength, and thereby interchanging field equations with Bianchi identities. 
In this context, we shall discuss  and generalise our results in the relativistic 
region and give our analysis in four dimensions in the subsequent sections.

In Section-\ref{gf-v}, we give the results of gauge fields and in 
Section-\ref{fermions}, we 
discuss the elementary fermions in detail. In Section-\ref{action}, we discuss the
action integral for supergravity, following Mishra~\cite{Mishra92} 
and Nilles~\cite{Niles}.
In this, we find that in the four dimension( not in ten dimension), one can most 
possibly write down a renormalisable Lagrangian. Our ideas will be somewhat different
than those of Mishra and Nilles. Because, they will reflect our motivation of 
including all aspects of the theory with our proposal for the four dimensional 
superstring.

We also discuss the renormalisability of the supergravity
Lagrangian in Section-\ref{renorm}. This appears renormalisable. In last 
Section-\ref{concl}, we give the reasons for having three generation in
the Standard Model. It is important therefore that there should be a 
renormalisable theory of strong, electromagnetic, weak and gravitational 
interactions at the same energy level.

\section{Superstring in Four Dimensions}\label{superstring}
In bosonic string theory, one starts with the  Nambu-Goto string
~\cite{Nambu70, Goto71}, given by action
\begin{equation}
S_B=-\frac{1}{2\pi}\int d^2\sigma
\left ( \partial_{\alpha}X^{\mu}\partial^{\alpha} X_{\mu} \right ),~~~\mu
=0,1,2,...,25,\label{e1}
\end{equation}
where $\partial_{\alpha}=(\partial_{\sigma},\partial_{\tau})$. The general
expression for the  energy momentum tensors at two world
sheet points z and $\omega$

\begin{equation}
2< T(z)T(\omega) > = \frac{C}{(z-\omega)^4} +\hdots.\label{e2}
\end{equation}
$C$,the coefficient of the most divergent term in the above equation (\ref{e2}),
is the central charge. Methods and principles of calculation of $C$, for a variety 
of strings, has been given in reference~\cite{Deo06}. For free bosons, the central
charge   $C_B~=~\delta_{\mu}^{\mu}$, with $\mu$=0,1,....25; thereby the string 
action makes sense only in 26 dimensions. Due to this feature, one has
to discard the theory and has to go to 10 dimensional supersting.

However, the 26-dimensional string theory can be made to work as a superstring
having four bosonic  and  forty four fermionic degrees of freedom with 
SO(44) symmetry. One uses the Mandelstam's proof of equivalence between one 
boson and two fermions in 1+1  field theory. This is true in 
a finite intervals or circles.

The 44 fermions can form 11 Lorentz vectors. The String action $S_{FB}$, including 
these fermions, can be written as
\begin{equation}
S_{FB}=-\frac{1}{2\pi}~\int~d^2 \sigma
\left [ \partial^{\alpha}X^{\mu}(\sigma,\tau)~
\partial_{\alpha}X_{\mu}(\sigma,\tau) -
i\sum_{j=1}^{11}\bar{\psi}^{\mu,j}
\rho^{\alpha}\partial _{\alpha}\psi_{\mu,j} \right ].\label{e3}
\end{equation}
We have
\begin{eqnarray}
\rho^0 =
\left (
\begin{array}{cc}
0 & -i\\
i & 0\\
\end{array}
\right ),~~~~~
\rho^1 =
\left (
\begin{array}{cc}
0 & i\\
i & 0\\
\end{array}
\right ),~~~~~\textnormal{and}~~~~
\bar{\psi}=\psi^{\dag}\rho^0.
\end{eqnarray}

Here $\rho^{\alpha}$'s are imaginary, so the Dirac operators
$\rho^{\alpha}\partial_{\alpha}$ are real. In
this representation of the Dirac algebra, the components of the world
sheet spinor $\psi^{\mu,j}$ are real and they are Majorana spinors.
But the action~(\ref{e3}) is not supersymmetric.
The eleven $\psi^{\mu,j}_A$ have to be further
divided into two species; $\psi^{\mu,j} ,~~ j=1,2,...,6$
and $\phi^{\mu, k},~~k=7,8,...,11$.
For the group of six, the positive and negative parts of
$\psi^{\mu,j}=\psi^{(+)\mu,j} +\psi^{(-)\mu,j}$,
whereas for the group of five, allowed the freedom of
phase of creation operators for Majorana fermions
in $\phi^{\mu,k} = \phi^{(+)\mu,k}-\phi^{(-)\mu,k}$.
This is due to the fact that we have Majorana like neutrinos rather than
Dirac like ones. In fact `neutrinos' are $6\times 4= 24$ in number and can be
taken as right handed.The left handed ones can be $5\times 4=20$ in
number~\cite{Mohapatra91}.
The action is now

\begin{equation}
S= -\frac{1}{2\pi} \int d^2\sigma
\left [\partial_{\alpha}X^{\mu}\partial^{\alpha}X_{\mu}
-i~\bar{\psi}^{\mu,j}~\rho^{\alpha}~\partial_{\alpha}~\psi_{\mu,j}
+ i~\bar{\phi}^{\mu,k}~\rho^{\alpha}~
\partial_{\alpha}~\phi_{\mu,k}\right ],
\label{e4}
\end{equation}
where j=1,2,...,6 and k=7,8,...,11.

Besides  the $SO(3,1)$, the action (\ref{e4}) is invariant
under $SO(6)\otimes SO(5)$. It is also
invariant under the transformations
\begin{eqnarray}
\delta X^{\mu} &=&\bar{\epsilon}~(e^j\psi^{\mu}_j - e^k\phi^{\mu}_k),\\
\delta\psi^{\mu,j}
&=& - ie^j\rho^{\alpha}\partial_{\alpha}X^{\mu}~\epsilon,\\
\textnormal{and}~~~~~\delta\phi^{\mu,k}
&=& ie^k\rho^{\alpha}\partial_{\alpha}X^{\mu}~\epsilon.\label{e5}
\end{eqnarray}
Here $\epsilon$ is a constant anticommuting spinor.
$e^j$ and $e^k$ are eleven numbers of a
row with i.e. $e^5$=(00001000000) and $e^{10}$=(00000000010) and 
$\sum_{j=1}^6 e^j e_j$=6 and $\sum_{k=7}^{11} e^k e_k$=5.
In the formulation of the theory, one must have the proof that the
commutator of two supersymmetric transformations gives a world sheet
translation. With $\Psi^\mu=e^j\psi^{\mu}_j - e^k\phi^{\mu}_k,$, 
which is the superpartner of $X^{\mu}$, we show that
\begin{equation}
\left ( \delta_1,\delta_2\right )X^\mu=a^{\alpha}\partial_\alpha X^\mu,\label{e6}
\end{equation}
and
\begin{equation}
\left ( \delta_1,\delta_2\right )\Psi^\mu=a^{\alpha}
\partial_\alpha \Psi^\mu,\label{e7}
\end{equation}
where $a^\mu=2i\bar{\epsilon}\rho^\alpha\epsilon_2$ as expected 
in supersymmetric theories.

We proceed to quantise the theory. Let b and b$'$ be the 
quanta of $\psi$ and $\phi$ fields in Neveu-Schwarz(NS) formulation 
and d and d$'$ are the quanta of the above
fields in Ramond(R) formulation. Specifically,for the X's, we have,
\begin{equation}
X^{\mu}(\sigma,\tau)= x^{\mu} +p^{\mu}\tau + 
i\sum_{n\neq0}\frac{1}{n} \alpha_n^{\mu}~
e^{-in\tau}~\cos(n\sigma). \label{e8}
\end{equation}
In terms of complex coordinates $z=\sigma + i\tau$ and
$\bar{z}=\sigma - i\tau$, we have,
\begin{equation}
X^{\mu}(z,\bar{z})= x^{\mu} -i\alpha^{\mu}_0\ln |z| 
+i\sum_{m\neq 0}\frac{1}{m} \alpha_m^{\mu}~z^{-m}. \label{e9}
\end{equation}
Further,
\begin{equation}
\psi_{\pm}^{\mu,j}(\sigma,\tau)=
\frac{1}{\sqrt{2}}\sum_{r\in Z+\frac{1}{2}}
b^{\mu,j}_r~e^{-ir(\sigma \pm\tau)},~~~~~
 and~~~~~\phi_{\pm}^{\mu,k}(\sigma,\tau)=
\frac{1}{\sqrt{2}}\sum_{r\in Z+\frac{1}{2}}
b^{'\mu,k}_r~e^{-ir(\sigma \pm\tau)}
~~~ \textnormal{for ~~~ NS~~ sector},\label{e10}
\end{equation}
and,
\begin{equation}
\psi_{\pm}^{\mu,j}(\sigma,\tau)=
\frac{1}{\sqrt{2}}\sum_{m=-\infty}^{\infty}
d^{\mu,j}_m~e^{-im(\sigma \pm\tau)},~~~~and
~~~~~\phi_{\pm}^{\mu,k}(\sigma,\tau)
=\frac{1}{\sqrt{2}}\sum_{m=-\infty}^{\infty}
d^{'\mu,k}_m~e^{-im(\sigma \pm\tau)}~~ 
\textnormal{for~~ R~~sector}.\label{eq11}
\end{equation}
In the formulation of superstring theory in the above section,
a principal role has been played by the proof that the commutator of 
two supersymmetric transformations gives a world
sheet translation. Therefore, it is necessary to have an
exact framework in which the super
Virasoro conditions can emerge as gauge conditions. For this,
the action given in (\ref{e4})
should incorporate the superconformal invariance of a full
superstring theory. These have been dealt with by one of us in Ref.\cite{Deo06}

Varying the field and Zweibein, the Noether current $J^{\alpha}$ and
the energy momentum tensor $T_{\alpha\beta}$ vanishes,
\begin{equation}
J_{\alpha}= \frac{\pi}{2e}\frac{\delta S}{\delta \chi^{\alpha}}
=\rho^{\beta}
\rho_{\alpha}\bar{\Psi}^{\mu}
\partial_{\beta}X_{\mu}=0,\label{e12}
\end{equation}
and
\begin{equation}
T_{\alpha\beta}=\partial_{\alpha}X^{\mu }
\partial_{\beta}X_{\mu }- 
\frac{i}{2}\bar{\Psi}^{\mu}\rho_{(\alpha}\partial_{\beta )}
\Psi_{\mu}=0.\label{e13}
\end{equation}
These are the super Virasoro constrain equations.
In a light cone basis, the vanishing of the lightcone
components are obtained from the variation of the action given in
equations (\ref{e12}) and (\ref{e13}),
\begin{equation}
J_{\pm}=\partial_{\pm}X_{\mu}\Psi^{\mu}_{\pm}=0,\label{e16a}
\end{equation}
and
\begin{equation}
T_{\pm\pm}=
\partial_{\pm}X^{\mu}\partial_{\pm}X_{\mu}
+\frac{i}{ 2}\psi^{\mu j}_{\pm}\partial_{\pm}
\psi_{\pm\mu,j }- 
\frac{i}{2}\phi_{\pm}^{\mu k}\partial_{\pm}\phi_{\pm\mu,k},\label{e15}
\end{equation}
where $\partial_{\pm}=\frac{1}{2}(\partial_{\tau} \pm\partial_{\sigma})$.

For a brief outline, let $L_m$, $G_r$ and $F_m$ be the Super
Virasoro generators of energy, momenta and currents. Then,
\begin{eqnarray}
L_m&=&\frac{1}{\pi}\int_{-\pi}^{\pi}d\sigma e^{im\sigma}
T_{++}\nonumber\\
&= &\frac{1}{2}\sum^{\infty}_{-\infty}:
\alpha_{-n}\cdot\alpha_{m+n}: +\frac{1}{2}
\sum_{r\in z+\frac{1}{2}}(r+\frac{1}{2}m): 
(b_{-r} \cdot b_{m+r} - b_{-r}' \cdot b_{m+r}'),~~~~\textnormal{for}~~~~~\textnormal{NS}
\nonumber\\
&=&\frac{1}{2}\sum^{\infty}_{-\infty}:\alpha_{-n}
\cdot\alpha_{m+n}: +\frac{1}{2}
\sum^{\infty}_{n=-\infty}(n+\frac{1}{2}m): 
(d_{-n} \cdot d_{m+n} - d_{-n}'
\cdot d_{m+n}'),~~~ ~\textnormal{for}~~~~~~\textnormal{R}\\
G_r &=&\frac{\sqrt{2}}{\pi}\int_{-\pi}^{\pi}d\sigma 
e^{ir\sigma}J_{+}=
\sum_{n=-\infty}^{\infty}\alpha_{-n}\cdot 
\left( e^jb_{n+r,j}- 
e^kb'_{n+r,k}\right ), ~~~~~~~~~~~~~~~~\textnormal{for}~~~~~~~~~\textnormal{NS}\\
\textnormal{and}\nonumber\\
F_m &=&\sum_{-\infty}^{\infty} \alpha_{-n}\cdot 
\left( e^jd_{n+m,j}- e^kd'_{n+m,k}\right ) =\sum_{-\infty}^{\infty}\alpha_{-n}\cdot
{\cal D}_{n+m}.~~~~~~~~~ 
~~~~~~~~~~~~\textnormal{for}~~~~~~~~~~~~~~~~ \textnormal{R}\label{e18}
\end{eqnarray}
and satisfy the super Virasoro algebra, with central
charge $C=26$ for the action of equation (\ref{e4}),
\begin{eqnarray}
\left [L_m , L_n\right ] 
& = &(m-n)L_{m+n} +\frac{C}{12}(m^3-m)\delta_{m,-n},\\
\left [L_m , G_r\right ] 
& = &(\frac{1}{2}m-r)G_{m+r}, 
~~~~~~~~~~~~~~~~~~~~~~~~~~~~~\textnormal{for}~~~~~~~~~~~~~\textnormal{NS}\\
\{G_r , G_s\} & =& 2L_{s+r} +\frac{C}{3}(r^2-
\frac{1}{4})\delta_{r,-s},\label{e19}\\
\left [L_m , F_n\right ] 
& = & (\frac{1}{2}m-n)F_{m+n},~~~~~~~~~~~~~~~~~~~~
~~~\textnormal{for}~~~~~~~~~~~~~~~~~~~~\textnormal{R}\\
\{F_m, F_n\} & = & 2L_{m+n} +\frac{C}{3}(m^2-1)
\delta_{m,-n},~~~\;\;\;\; m\neq 0.\label{e19a}
\end{eqnarray}
Equations (\ref{e19}) and (\ref{e19a}) can be
obtained using Jacobi identity.

This is also known that the normal ordering constant of
$L_o$ is equal to one and we define the physical
states, satisfying
\begin{eqnarray}
(L_o-1)|\phi>&=&0,~~~L_m|\phi>=0,~~~G_r|\phi>
=0~~~ \text{for}~~~~ r,m>0,\;\;\;:\textnormal{NS}\;\;\;\textnormal{Bosonic}\label{e20}\\
 L_m|\psi>&=&F_m|\psi>=0,\;\;\;\;\;\;\;\;\;\;\;\;\;\;\;
\textnormal{for}\;\;\;\;\; m>0,\;\;\;\;\;\;\;\; ~~~~~~~~~~:\textnormal{R}\;\;\;\;
\textnormal{Fermionic}\label{e21}\\
\textnormal{and}~~~(L_o-1)|\psi>_{\alpha}&=&(F_o^2-1)|\psi>_{\alpha}=0.\label{e22}
\end{eqnarray}
So we have,
\begin{equation}
(F_o +1)|\psi_+>_{\alpha}=0\;\;\;\; 
\textnormal{and}\;\;\; (F_o-1)|\psi_->_{\alpha}=0.~~~~~~~~~~~~~:\textnormal{R}.\label{e23}
\end{equation}
These conditions shall make the string model ghost free.
It can be seen in a very simple way. Applying $L_o$ condition,
the state  $\alpha_{-1}^{\mu}|0,k>$ is massless.
The $L_1$ constraint gives the
Lorentz condition $k^{\mu}|0,k>=0$, implying a transverse
photon and with  $\alpha_{-1}^0|\phi>=0$ as Gupta-Bleuler condition.
Applying $L_2,~~L_3~~ ....$, constraints,
one obtains $\alpha_{m}^0|\phi>=0$. Further,
since $[\alpha_{-1}^0 , G_{r+1}]|\phi>=0$,~~
we have $b_{r,j}^0|\phi>=0$ and $b_{r,k}^{'0}|\phi>=0$.
All the time components are eliminated from Fock space.
                                                          
Ghosts do not couple to physical states, therefore these states
must be of the form (up to null states)\cite{Friedman86}
\begin{equation}
|\{m\}p\rangle_M\otimes~ c_1|0\rangle_G,
\end{equation}
where $|\{m\}p\rangle_M$ denotes the occupation numbers and momentum of
physical matter states. The operator $c_1$ lowers the energy of 
the states by one unit and is necessary for BSRT invariance. The ghost 
is responsible for lowering the ground state energy producing shiftable 
tachyon($F_2$ picture)
\begin{equation}(
L_0^M-1)|phys\rangle = 0.
\end{equation}
Therefore, the mass shell condition is
\begin{eqnarray}
\alpha'M^2&=&N^B+N_{NS}^F -1~:~~~~~~~~~\textnormal{NS},\\
\alpha'M^2&=&N^B+N_{R}^F -1~~~:~~~~~~~~~\textnormal{R},
\end{eqnarray}
where
\begin{eqnarray}
N^B&=&\sum_{m=1}^\infty \alpha_{-m}\alpha_m\\
N^F_{NS}&=&\sum_{r=1/2}^\infty r(b_r\cdot b_r +b'_r\cdot b'_r)~~~~~:~~~~\textnormal{NS},
\end{eqnarray}
and
\begin{equation}
N^F_{R}=\sum_{m=1}^\infty m(d_m\cdot d_m +d'_m\cdot d'_m)~~~~~:~~~~\textnormal{R}.
\end{equation}

In general,
\begin{equation}
\alpha'M^2=-1,-1/2,0,1/2,1,3/2,.....,
\end{equation}
in the NS sector. This,
in the shifted Hilbert state, is
\begin{equation}
\alpha'M^2=-1/2,0,1/2,1,3/2,.....
\end{equation}
Due to presence of Ramond and Neveu-Schwartz sectors, with periodic and anti-
periodic boundary conditions, we can effect a GSO projection\cite{Gliozzi76}
on the mass spectrum on NSR model\cite{Kaku99}. The projection, as desired, is
\begin{equation}
G=\frac{1}{2}\left ( 1+ (-1)^{F+F'}\right ),
\end{equation}
where $F=\sum b_r\cdot b_r$ and $F'=b'_r\cdot b'_r$. This will
eliminate the half integral value from the mass spectrum by choosing G=1
including the tachyon at $\alpha M^2=-\frac{1}{2}$.

We have, therefore
\begin{equation}
\alpha'M^2= -1, 0 , 1, 2, 3,~~~~~~\textnormal{for}~~~~~~:R.
\end{equation}
The G.S.O. projection eliminates the half integral values.
The tachyonic self energy of
bosonic sector $<0|(L_o-1)^{-1}|0>$ is cancelled
by $-<0|(F_o+1)^{-1}(F_o-1)^{-1}|0>_R$,
the negative sign being due to the fermionic loop.
Such tadpole cancellations have been
noted also by Chattaraputi et al in reference~\cite{Chattaraputi021}.
One can proceed a step further and write down the world sheet charge.
To be more sure, consider the supersymmetric charge
\begin{equation}
Q=\frac{1}{\pi}\int_0^\pi\rho^o\rho^{\alpha\dag}\partial_\alpha X^\mu
\Psi_\mu d\sigma .
\end{equation}
The supersymmetric result is
\begin{equation}
\sum\{Q_\alpha^\dag, Q_\alpha\}=2H~~~ and ~~~ \sum_\alpha|Q_\alpha|\phi_o>|^2=
2<\phi_o|H|\phi_o>.
\end{equation}
The ground state is of zero energy.There is no tachyon. The physical mass
spectrum in both sectors are integral numbers of Regge trajectory,
$\alpha M^2=0,1,2,.....$

\section{Polchinski Superstring}\label{polch}                                                           
The last section-\ref{superstring} summarises what have been done earlier. 
Now we propose to construct a theory in four dimension within the following 
observation of Polchinski~
\cite{Polchinski98}, with a link between superstring states and generator
in Conformal Field Theory(CFT).
For example, for any conserved charge Q, the operator equation QA is 
Q($\psi_A$). If A is a unit operator $\hat{\mathbb {I}}$ and 
\begin{equation}
Q=\alpha_m= (2\pi)^{-1}\oint dt~z^m~\partial X ~~~~ \textnormal{for}~ m~ >~0,  
\end{equation}
so that $\partial X$ is 
analytic and the integral vanishes for $m\geq 0$. We get $\alpha_m|\psi_I>=$0,
for m $\geq$ 0. The exact correspondance of the unit operator and string 
operators $|0,0>$ is thus
\begin{equation}
\hat{\mathbb{I}} \leftrightarrow |0,0>.
\end{equation}
Similarly, 
\begin{equation}
:e^{k.X(z)}:\leftrightarrow |0,k>.
\end{equation}
Unitarity gives the normalisation                                                    
\begin{equation}
<0,k'|0,k>=2\pi \delta(k-k').
\end{equation}
This can be generalised  to momentum states, with $k_0=\vec{k}$,
\begin{equation}
<0,k'|0,k>=(2\pi)^3(2k_0) \delta^{(3)}(k-k').
\end{equation}

With these remarks, we shall follow the work of Li~\cite{Li74},
where  $\frac{1}{2}n(n-1)$ generators of O(n) are represented by 

\begin{equation}
L_{ij}= X_i\frac{\partial}{\partial X_j}-X_j
\frac{\partial}{\partial X_i}, ~~i,j=1,....,n.
\end{equation}
The commutation among the generators, called Lie algebra
can be worked out. The rule is to write
\begin{equation}
\left [ \frac{\partial}{\partial X_i}, X_j\right] =\delta_{ij},
\end{equation}
so that we have 
\begin{equation}
\left [ L_{ij}, L_{kl} \right ]
= \delta_{jk} L_{il}+\delta_{il} L_{jk}-
\delta_{ik} L_{jl}-\delta_{jl} L_{ik}.
\end{equation}
Hence one must have $\frac{1}{2}n(n-1)$ vector gauge
bosons $W^{\mu}_{ij}$ with the transformation law
\begin{equation}
W^{\mu}_{ij}\rightarrow W^{\mu}_{ij}+\epsilon_{ik}
W^{\mu}_{kj}+ \epsilon_{jl}  W^{\mu}_{li}, ~~~~
W^{\mu}_{ij}= -W^{\mu}_{ji},
\end{equation}
where $\epsilon_{ij}= - \epsilon_{ji}$ is the infinitesimal
parameters which characterise such rotation in
$O(n)$. Under gauge transformation of the second kind, we have
\begin{equation}
W^{\mu}_{ij}\rightarrow W^{\mu}_{ij}+\epsilon_{ik}
W^{\mu}_{kj}+ \epsilon_{jl}  W^{\mu}_{li}+\frac{1}{g}                           
\partial^{\mu}\epsilon_{ij}.
\end{equation}
The Yang-Mills Lagrangian is then written as
\begin{equation}
L=-\frac{1}{4}|F^{\mu\nu}_{ij}|^2,
\end{equation}
with
\begin{equation}
F^{\mu\nu}_{ij} = \partial^{\mu}W^{\nu}_{ij}-
\partial^{\nu}W^{\mu}_{ij}+g\left (W^{\mu}_{ik}W^{\nu}_{kj}
-W^{\nu}_{ik}W^{\mu}_{kj}\right ),
\end{equation}
$F^{\mu\nu}_{ij}$ has the obvious properties, namely,
\begin{equation}
\Box F^{\mu\nu}_{ij}=0,~~~~ \partial_{\mu}
F^{\mu\nu}_{ij}=\partial_{\nu}F^{\mu\nu}_{ij}=0,
~~~and~~~~~~~ F^{\mu\mu}_{ij}=0.
\end{equation}
There are two sets of field strength tensors
which are found in the model, one for $SO(6)$ and the 
other for $SO(5)$. Since $\Box F^{\mu\nu}_{ij}=0 $, one can
take plane wave solution and write
\begin{equation}
F^{\mu\nu}_{ij}(x)=F^{\mu\nu}_{ij}(p)~e^{ipx},
\end{equation}
so that,
\begin{equation}
p^2 F^{\mu\nu}_{ij}(p)=p_{\mu}F^{\mu\nu}_{ij}(p)
=p_{\nu}F^{\mu\nu}_{ij}(p)=0,~~~
F^{\nu\nu}_{ij}(p)=0~~~~{and}~~~~~
F^{\mu\nu}_{ij}(p)=-F^{\mu\nu}_{ji}(p).
\end{equation}
These are physical state conditions (\ref{e20})-(\ref{e22}) as well and
\begin{equation}
L_0 F^{\mu\nu}_{ij}(p)=0,~~~~~~~~G_{\frac{1}{2}} F^{\mu\nu}_{ij}(p)=0,
~~~~~{and}~~~~~~L_1 F^{\mu\nu}_{ij}(p)=0.\label{e71}
\end{equation}
The field strength tensor, satisfying (\ref{e71}), is found to be
\begin{equation}
F^{\mu\nu}_{ij}(p)= b^{\mu\dag}_i~b^{\nu\dag}_j |0,p> 
+ \epsilon_{ij}(p^{\mu}\alpha_{-1}^{\nu}-p^{\nu}
\alpha_{-1}^{\mu})|0,p>,\label{e71a}
\end{equation}
with $\epsilon_{ij} = e_i^{\alpha} e_j^{\beta} 
\varepsilon_{\alpha\beta},
~~(i,j)$=1,...,6~for O(6) and 1,...,5 for O(5)
with $b'$ replaced by $b$.
For simplicity, we drop the $\dag$'s.
In terms of the excitation quanta of the string,
the vector generators are
\begin{equation}
W^{\mu}_{ij}=\frac{1}{\sqrt{2ng}}~n_{\kappa}
\epsilon^{\kappa\mu\nu\sigma}b_{\nu,i} b_{\sigma,j} 
+ \epsilon_{ij}\alpha_{-1}^{\mu},
\end{equation}
where $n_{\kappa}$ is the time-like four vector
and can be taken as (1,0,0,0). One finds that
\begin{eqnarray}
\partial^{\mu}W^{\nu}_{ij}-\partial^{\nu}W^{\mu}_{ij}
&=&p^{\mu}W^{\nu}_{ij}-p^{\nu}W^{\mu}_{ij}\nonumber\\
&=&\frac{1}{\sqrt{2ng}}\left ( n_{\kappa}
\epsilon^{\kappa\nu\lambda\sigma}p^{\mu}
-n_{\kappa}\epsilon^{\kappa\mu\lambda\sigma}
p^{\nu}\right )b_{\lambda,i} b_{\sigma,j} 
+ \epsilon_{ij}(p^{\mu}\alpha_{-1}^{\nu} 
- p^{\nu}\alpha^{\mu}_{-1}).
\end{eqnarray}
As $\mu$ must be equal to $\nu$, if $\kappa,~\lambda ~
\textnormal{and}~ \sigma$ are the same, then the first term vanishes 
and we have
\begin{eqnarray}
g\left (W^{\mu}_{ik}W^{\nu}_{kj} 
- W^{\nu}_{ik}W^{\mu}_{kj}\right )
&=&\frac{1}{2n}\left (  n_{\kappa}~
\epsilon^{\kappa\mu\lambda\sigma}~
b_{\lambda,i} b_{\sigma,k}
~n_{\kappa'}~\epsilon^{\kappa'\nu\lambda'\sigma'}~
b_{\lambda',k} b_{\sigma,j}
-~~~\mu\leftrightarrow \nu\right )
+ \epsilon_{ik}\epsilon_{kj}\left [\alpha_{-1}^{\mu}, 
\alpha_{-1}^{\nu}\right ]\nonumber\\
&=&b_i^{\mu}b_j^{\nu}.
\end{eqnarray}
In the above, we have used
\begin{equation}
\left \{ b_{\lambda'k}, b_{\sigma k}\right \} =
\eta_{\lambda'\sigma}~\delta_{kk}=~n~\eta_{\lambda'\sigma},
\end{equation}
and the creation operators $\alpha_{-1}^{\mu}$ and $ \alpha_{-1}^{\nu}$ commute.
Equation (\ref{e71a}) is referred as the field strength.
Since the product of pairs of $b$ and $b'$ commute, the gauge group of
the action~(\ref{e4}) is the product group $SO(6)\otimes SO(5)$.
This is same as the symmetry group of the action (\ref{e4}).
For O(6), we have i,j=1,........,6 and for O(5), we have to replace b by b' 
with~ i,j=1,....,5. Now, we denote 
\begin{equation}
F_{\mu\nu}^{(\alpha)}=\partial_\mu A_\nu^{(\alpha)}-
\partial_\nu A_\mu^{(\alpha)},\label{eq:71aa}
\end{equation}
as a 12-dimensional vector which can be found out from (\ref{e71a}) and that
\begin{eqnarray}
F^{\mu\nu(\alpha)}(x)&=&\int d^3x\frac{1}{\sqrt{2p^0}}\epsilon^{\alpha ij}
(b_{i}^{\mu\dag }b_j^{\nu\dag})|0,p>e^{ipx},~~~~~~\alpha= 2,...,7\label{eq:71ab}\\
&=&\int d^3x\frac{1}{\sqrt{2p^0}}\epsilon^{\alpha ij}
(b_{i}^{'\mu\dag }b_j^{'\nu\dag})|0,p>e^{ipx},~~~~\alpha= 8,...,12,\label{eq:71ac}\\
\textnormal{and}~~~~~~~~~~~~~~~~~~~F^{\mu\nu(1)}(x)&=&\int d^3x\frac{1}{\sqrt{2p^0}}
(p^{\mu}\alpha_{-1}^{\nu}-p^{\nu}\alpha_{-1}^{\mu})|0,p>e^{ipx}~~~~
\alpha = 1.\label{eq:71ad}
\end{eqnarray}
It should be noted that we have not taken account of the degrees of 
freedom arising out of SO(3,1) vectors. We shall do what follows, by means of a
complex operator field $\lambda=\lambda_1\pm i\lambda_2$.

\section{Invariance and Duality}\label{invariant}
The basic idea is to prove that the fields we have observed are dual~
\cite{Schwarz94,Jn92,Ashok08} . 
With the  pair of field variables $ E^{(\beta) i}~~and~~ B^{(\alpha) i}$, 
the action S, in flat spacetime,
\begin{equation}
S=\frac{1}{2}\int d^4x\left (B^{(\alpha) i}{\cal L}_{\alpha\beta} E^{(\beta) i}
+ B^{(\alpha) i}B^{(\alpha) i}\right ),\label{e70}
\end{equation}
where
\begin{equation}
E^{\alpha}_{ i}=\partial_0 A_i^{(\alpha)}-
\partial_i A_0^{(\alpha)},~~~B^{(\alpha) i}=\epsilon^{ijk}\partial_j 
A_k^{(\alpha)},~~1\leq i,j,k\leq 3,\label{eq:70}
\end{equation}
with
\begin{eqnarray}
{\cal L} =
\left (
\begin{array}{cc}
0 & 1\\
-1 & 0\\
\end{array}
\right ).
\end{eqnarray}
In defining (\ref{eq:70}), we have used (\ref{eq:71aa}).
The  action S in ~(\ref{e70}) is invariant under the following transformations
\begin{equation}
\delta A_0^{(\alpha)}=\Psi^{(\alpha)},~~~~~\textnormal{and}~~~~~~
\delta A_i^{(\alpha)}=\partial_i \Lambda^{(\alpha)}.
\end{equation}
For the $ \Psi^{(\alpha)}$, we can set $ A_0^{(\alpha)}$=0. The equation of motion
 of the field  $A_i^{(2)}$ is
\begin{equation}
\epsilon^{ijk}\partial_i(~B^{(j)k}-E^{(j)}_k)=0.
\end{equation}
Since no time derivative is involve, $A_i^{(2)}$ can be treated  
as an auxiliary field. We can eliminate this so that we are left with 
\begin{equation}
B^{(2)k}=E_k^{(1)}+\partial_k\phi,
\end{equation}
for some $\phi$ which we shall take to be zero so that we have 
\begin{equation}
 B^{(2)k}=E_k^{(1)},\label{e71b}
\end{equation}
and on substitution, the action becomes
\begin{equation}
S=-\frac{1}{2}\int d^4x (~B^{(1)i}B^{(1)i}-E^{(1)}_iE^{(1)}_i),\label{e72}
\end{equation}
in the gauge $A_0^{(1)}$=0. The action in equation (\ref{e70}) is manifestly
invariant under duality symmetry
\begin{equation}
A_\mu^{(\alpha)}\rightarrow {\cal L}_{\alpha\beta}A_\mu^{(\beta)},
\end{equation}
which implies the transformation
\begin{equation}
\left (
\begin{array}{c}
B^{(1)i}\\
E^{(1)}_i\\
\end{array}
\right )={\cal L}
\left (
\begin{array}{c}
B^{(1)i}\\
E^{(1)}_i\\
\end{array}
\right ),
\end{equation}
when we use the equation of motion for (\ref{e71b}) of $A_i^{(2)}$.
The action (\ref{e1}) becomes much simpler in curved space time  for the field 
$A_\mu^{(1)}$,
\begin{equation}
S=-\frac{1}{4}\int d^4x \sqrt{-g}g^{\mu\rho}g^{\nu\sigma}F^{(1)}_{\mu\nu}
F^{(1)}_{\rho\sigma}.
\end{equation}
This can be quantised in the form
\begin{equation}
S=-\frac{1}{2}\int d^4x \left ( B^{(\alpha) i}{\cal L}_{\alpha\beta} E^{(\beta)}_ i
-\frac{g_{ij}}{\sqrt{-gg^{00}}}B^{(\alpha) i} B^{(\alpha) j}+
\epsilon_{ijk}\frac{g^{0i}}{g^{00}} B^{(\alpha) j}{\cal L}_{\alpha\beta} B^{(\beta) k}
\right ).\label{eq:73}
\end{equation}
We now generalise the above (\ref{eq:73}) to our case and get
\begin{equation}
S_A=\int d^4x \sqrt{-g}\left (
R-\frac{1}{{2\lambda_2^2}}\partial_\mu\lambda\partial^\mu\bar{\lambda}
-\frac{1}{4}\lambda_2F^{(a)}_{\mu\nu}(LML)_{ab}F^{(a)\mu\nu}+
\frac{1}{4}\lambda_2F^{(a)}_{\mu\nu}L_{ab}F^{(b)\mu\nu}
+\frac{1}{8}g^{\mu\nu} Tr(\partial_\mu ML\partial_\nu ML)
\right ).
\end{equation}
Here $\lambda =\lambda_1 \pm i\lambda_2$ and $A^{(a)}_\mu,~~3\leq a\leq~12$, are 10 
abelian gauge fields i.e.
\begin{equation}
F^{(a)}_{\mu\nu}=\partial_\mu A^{(a)}_\nu -\partial_\nu A^{(a)}_\mu,~~~ \tilde{F}^{(a)\mu\nu}
=\frac{1}{2}\left (\sqrt{-g}\right )^{-1}\epsilon^{\mu\nu\rho\sigma}F^{(a)}_{\rho\sigma},
\end{equation}
and 
\begin{eqnarray}
L=
\left (
\begin{array}{cc}
0&I_6\\
I_6&0\\
\end{array}
\right ).
\end{eqnarray}
$I_n$ is a $n \times n$ unitary matrix,
M is the $12\times12$ matrix valued scalar fields satisfying the constraint,
\begin{equation}
M^T=M,~~~~~~~~~\textnormal{and}~~~~~~~~~~~~M^TLM=L, 
\end{equation}
and is given by
\begin{eqnarray}
M=
\left (
\begin{array}{cc}
G^{-1}&G^{-1}B\\
-BG^{-1}&G-BG^{-1}B\\
\end{array}
\right ).
\end{eqnarray}
The internal metric $G_{\alpha\beta}=E^{(a)}_{\alpha}\delta_{ab}E_\beta^{(b)}$ and 
($G_{\alpha \mu}B_{\mu\alpha})$ have limit 0$\leq (\mu,\nu)\leq $3 and 
1$\leq$ a$\leq$ 6
and  the transformation can be realised as transformation of G and B.
$B_{\mu\nu}$ is determined from the equation
\begin{equation}
 H_{\mu\nu\rho}=\partial_\mu B_{\nu\rho}+A_\mu^{(a)}{\cal L}_{ab}F_{\nu\rho}^{(b)}=
-(\sqrt{-g})^{-1}\left (\frac{i}{\lambda_2^2}\right )~\epsilon_{\mu\nu\rho\sigma}
~\partial^\sigma\lambda_1.
\end{equation}
$H_{\mu\nu\rho}$ is the tensor which gives rise to pseudorensor $B_{\mu\rho}$.
The equation of motion have the symmetry
\begin{equation}
\lambda \rightarrow \frac{a\lambda+b}{c\lambda+d} ~~:~~~ F^{(a)}_{\mu\nu}
\rightarrow  c\lambda_2(ML)_{ab}\tilde{F}^{(b)}_{\mu\nu}+(c\lambda_1+d)F^{(a)}_{\mu\nu},
\end{equation}
with
\begin{equation}
ad-bc=1.
\end{equation}
Particularly, if a=0, b=1, c=1 and d=0, then the transformations
are
\begin{equation}
\lambda \rightarrow -\frac{1}{\lambda},~~F^{(a)}_{\mu\nu}\leftrightarrow -\lambda_1
~F^{(a)}_{\mu\nu}-\lambda_2(ML)_{ab})\tilde{F}^{(b)}_{\mu\nu}.
\end{equation}
If $\lambda_1$ =0, then the transformation takes electric field to 
magnetic field and vice versa. It also takes $\lambda_2$ to $\frac{1}{\lambda_2}$
i.e. this duality transformation takes a strong coupling to a weak coupling theory
and vice versa. The transformation is referred to as the strong-weak coupling or
electric-magnetic duality transformation.

We further note that the Ricci tensor  is
\begin{equation}
R_{\mu\nu}=\frac{\partial_\mu\bar{\lambda}\partial_\nu\lambda +
\partial_\nu\bar{\lambda}\partial_\mu\lambda}{4(\lambda_2)^2}+
2\lambda_2F^{(a)}_{\mu\rho}(LML)_{ab}F^{(b)\rho}_{\nu}-
\frac{1}{2}\lambda_2g_{\mu\nu} F^{(a)}_{\rho\sigma}(LML)_{ab}F^{(b)\rho\sigma},
\end{equation}
\begin{equation}
D_\mu\lambda-(ML)_{ab}F^{(b)\mu\nu}+\lambda_1\tilde{F}^{(a)\mu\nu}=0,
\end{equation}
\begin{equation}
\frac{D^\mu D_\mu\lambda}{(\lambda_2)^2}
+i\frac{D^\mu\lambda D_\mu\lambda}{(\lambda_2)^3}
-iF^{(a)}_{\mu\nu}(LML)_{ab}F^{(b)\mu\nu}+\tilde{F}^{(a)}_{\mu\nu}(L)_{ab}F^{(b)\mu\nu}
=0,\label{eq:77}
\end{equation}
where $D_\mu$ is the standard covariant derivative with metric $g^{\mu\nu}$.
The Bianchi identity of the field strength tensor $F^{(a)}_{\mu\nu}$ are given as
\begin{equation}
D_\mu\tilde{F}^{(a)\mu\nu}=0.\label{du}
\end{equation}
This complete our discussion on this aspect of the theory.
The last equation (\ref{du}) is due to duality.

\section{Gauge Fields}\label{gf-v}

Whether in supersymmetry or supergravity, there will be need for gauge fields.
The total number of these depend on the particular case under consideration. 
In our case, the hypercharge vector $B_\mu$ is given by the state
\begin{equation}
|\phi(p)> = \alpha_{-1}^{\mu}|0,p>B_{\mu}.
\end{equation}
$L_0$ condition fixes the mass of $B_\mu$ as zero and $L_1$ satisfy the Lorentz condition
$p^\mu B_\mu(x)=0$. We have also tensors $b^{\mu}_{-\frac{1}{2},i} ~b^{\mu}_{-\frac{1}{2},j}$
which are massless and they have vector fields $A_{ij}^\mu$ in terms of which
\begin{eqnarray}
G^{\mu\nu}_{ij}&=&\int\frac{ d^3p}{\sqrt{2p_0}}~e^{ip\cdot x}\left ( 
b^{\mu}_i b^{\nu}_j -  b^{\nu}_i b^{\mu}_j -
\frac{2}{3}\delta_{ij} b^{\mu}_l b^{\nu}_l\right )|0,p>\\
&=& \partial^{\nu}A^{\mu}_{ij} - \partial^{\mu}A^{\nu}_{ij} +\left ( A^{\mu}_{il}
A^{\nu}_{jl}- A^{\nu}_{il}A^{\mu}_{jl}\right ).
\end{eqnarray}
Here   $i,j$=1,2...,6 and we have omitted -$\frac{1}{2}$.    
The eight gluons are obtained by using the eight Gell-Mann ~$\lambda_l$~matrices.
\begin{equation}
V_l^{\mu}=(\lambda_l)_{ij}A^{\mu}_{ij}.\label{v1}
\end{equation}
Similarly for $b'$, we get the $W^{\mu}$-mesons
\begin{eqnarray}
W^{\mu\nu}_{ij}&=&\int\frac{ d^3p}{\sqrt{2p_0}}~e^{ip\cdot x}
 \left (b_i^{'\mu} b_j^{'\nu} - b_i^{'\nu} b_j^{'\mu} - \delta_{ij}
b_l^{'\nu} b_l^{'\mu}\right )|o,p>,\\
&=& \partial^{\nu}W^{\mu}_{ij} - \partial^{\mu}W^{\nu}_{ij} + \left ( W^{\mu}_{il}
W^{\nu}_{jl}- W^{\nu}_{il}W^{\mu}_{jl}\right ),
\end{eqnarray}
with
\begin{equation}
W^{\mu}_l = ({\tau_l})_{ij}W^{\mu}_{ij},\label{v2}
\end{equation}
We have got $i,j$=1,2...,5.
Here $\tau_l$, $l$=1,2,3 are the $2 \times 2$ isospin matrices.
Thus  there are 8 gluons and 3 W-bosons and one hyperchage vector $B_\mu$. All 
together, we have 12 vector mesons. We call all of them as the componets 
of $W^a_{\mu\nu}$.  However, these vector mesons should be the 
part of the standard model group $SU_C(3)\otimes SU_L(2)\otimes U_Y(1)$.
Without breaking the symmetry,, one can take the other method where we take the 
Wilson loop
\begin{equation}
U_{\gamma}= P~exp\left (\oint_{\gamma} A_{\mu}\;dx^{\mu}\right ).
\end{equation}
$P$ represents the ordering of each term with respect
to the closed path $\gamma$.  This breaking can be
accomplished by one element $U_0$ of SU(4), such that
\begin{equation}
U_0^2=1,
\end{equation}
with SO(6)=SU(4), so that the descendant groups are  
\[SU_C(3)\otimes U_{B-L}(1)\otimes Z_2,\] 
$Z_2$ being the permutation group. Similarly,
$SO(5)\rightarrow SO(3)\otimes SO(2)=SU(2)\otimes U(1)$.
We have,
\begin{equation}
\frac{SO(5)}{Z_2}=SU(2)\otimes U(1).\label{e80a}
\end{equation}
Thus
\begin{equation}
\frac{SO(6) \otimes SO(5)}{Z_2 \otimes Z_2}=
SU_C(3)\otimes U_{B-L}(1)\otimes U_R(1) \otimes SU_L(2),
\end{equation}
making an identification with the usual low energy phenomenology.
But this is not the  standard model. We have an additional U(1).
However, there is an instance in $E_6$, where there
is a reduction of rank by one and several U(1)'s.
Following the same idea, we may take
\begin{equation}
U_{\gamma}=(\alpha_{\gamma}) \otimes 
\left(
\begin{array}{ccc}
\beta_{\gamma}&&\\&\beta_{\gamma}&\\&&
\beta^{-2}_{\gamma}\end{array}
\right) 
\otimes 
\left( 
\begin{array}{ccc}
\delta_{\gamma} &&\\&
\delta^{-1}_{\gamma}&\\&&
\end{array}
\right).
\end{equation}
$\alpha_{\gamma}^3$ =1 such that $\alpha_{\gamma}$ is the cube
root of unity. The structure of U inevitably lowers the rank by one and  we get,
\begin{equation}
\frac{SO(6)\otimes SO(5)}{Z_3}=SU_C(3)\otimes SU_L(2)\otimes  U_Y(1),
\end{equation}
which is our supersymmetric standard model.

To use this idea, let the color vector fields denoted by $V^l_{\mu}$, 
$l$=1,..,8, are gluon 
fields. $l$=9 is the $U_Y(1)$
field and $l$=10,11,12  stand for W-mesons fields. We shall use 
the temporal gauge~\cite{Christ80},
where $V^l_0$ =0. For each $V^l_i$, there are electric 
field strength $E_i^l$ and
magnetic field strength  $B_i^l$. Following Nambu~\cite{Nambu91}, the combination
\begin{equation}
\mathcal{F}_i^l=\frac{1}{\sqrt{2}}\left [ E_i^l +  B_i^l\right ],\label{eq90}
\end{equation}
satisfies the only nonvanishing equal time commutation relation
\begin{equation}
\left [ \mathcal{F}_i^{\dagger l}(x),  \mathcal{F}_j^{ m}(y)\right ] =i 
\delta^{lm}\epsilon_{ijk} \partial^k(x-y).\label{eq91}
\end{equation}
We then construct Wilson's loop line integrals to convert the ordinary 
derivatives acting on fermion fields to respective gauge covariant derivatives.
In the above, we note that the equations (\ref{eq90}) and (\ref{eq91}) 
are obtained if $E_i^l\rightarrow -B_i^l$ and $B_i^l\rightarrow E_i^l$

\section{The Fermions}\label{fermions}

Now we consider the fermionic sector. Let  $\Gamma_\mu$ represent the Dirac matrices 
and let $u(p)$ is the fermionic wave function with ~
$\Gamma^{\mu}d_{-1,j,\mu}|0,p>u(p)$~ and
~$\Gamma^{\mu}d'_{-1,k,\mu}|0,p>u(p)$~ are all massless states. Broadly j=1,..,6
are the colour sector and k=7,..,11 are the electroweak sector.
The current generator condition gives ~$\Gamma^{\mu}p_{\mu}u(p)=0$,~ the Dirac
equation for each of them. We have just that $SO(6)\otimes SO(5)\rightarrow Z_3
\otimes SU_C(3)\otimes SU_L(2)\otimes U_Y(1)$ and we have eleven 
fermions of zero mass. We
must find a method of putting all the fermions in the electroweak and 
color groups in one
generation. Choosing any four d's, we can write
\begin{equation}
\left\{d_i, d_j\right\}=2\delta_{ij},
\end{equation}
omitting the suffix -1, and prefix primes and supply a
factor 2

Since SO(4) can come from either SO(6) or SO(5)~\cite{Li74}. We can choose any
four of them and define new operators b's as follows
\begin{equation}
b_1=\frac{1}{2} (d_1 + i d_2),~~~~~b_1^*=\frac{1}{2} (d_1 - i d_2),~~~~~~~
b_2=\frac{1}{2} (d_3 + i d_4),~~~~and~~~~b_2^*=\frac{1}{2} (d_3 - i d_4).
\end{equation}
They satisfy the algebra ($i,j$=1,2),
\begin{equation}
\left\{b_i, b_j\right\}=0,~~~~~\left\{b^*_i, b^*_j\right\}=0~~~and ~~~
\left\{b_i, b^*_j\right\}=  2\delta_{ij}.
\end{equation}
This U(2) group is identified here as the isospin group of the strong color sector.We
can further define 
\begin{equation}
b_3=\frac{1}{2} (d_5 + i d_6),~~~~and~~~~b_3^*=\frac{1}{2} (d_5 - i d_6).
\end{equation}
Then we give the necessary assignments of all observed fermions in Table-\ref{tab:table1}.

\begin{center}
\begin{table}[h]
\caption{\label{tab:table1}}
{Particle Spectra, States and Quantum Numbers (Standard Model)}
\begin{tabular}{|c|c|c|c|c|c|}\hline
Particles &  States & $I_3$ &  Y  &  Q& States Assigned\\ \hline
$e_R$ & $\Gamma\cdot d^{'}_7|0>_{\alpha}$ &0 & -2&-1&$\psi_0$\\\hline
$ \left(
\begin{array}{c}
\nu_L \\ e_L \\ 
\end{array}  \right)
$
&
$
\left ( \begin{array}{c}
 \frac{1}{\sqrt{2}} \Gamma\cdot (d'_8 + i d'_9)|0>_{\alpha}\\
 \frac{1}{\sqrt{2}} \Gamma\cdot (d'_{10} + id'_{11}) |0>_{\alpha} \\
\end{array} \right )     
$
&
$
\begin{array}{c}
1/2 \\
-1/2 \\
\end{array}
$  
&
$
\begin{array}{c}
-1 \\
-1 \\
\end{array}
$
&
$
\begin{array}{c}
0 \\-1 \\ 
\end{array} $ 
&$(\psi_1,\psi_2)$\\
\hline
$u^{a,b,c}_R$  & $\Gamma\cdot d_{1,2,3}|0>_{\alpha}$ & 0 & 4/3 &2/3&
$(\psi_4,\psi_5,\psi_6)$\\ \hline
$d^{a,b,c}_R$  & $\Gamma\cdot d_{4,5,6}|0>_{\alpha}$ & 0 & -2/3 &-1/3&
$(\psi_7,\psi_8,\psi_9)$\\ \hline
$ \left (
\begin{array}{c}
u^{a,b,c}_L\\
d^{a,b,c}_L\\
\end{array}
\right )
$
&
$\left (
\begin{array}{c}
 \Gamma\cdot b_1|0>_{\alpha},
~~\Gamma\cdot b_1^*|0>_{\alpha}\\
~~ \Gamma\cdot b_2|0>_{\alpha},
~~\Gamma\cdot b_2^*|0>_{\alpha} \\
~~ \Gamma\cdot b_3|0>_{\alpha}, 
~~ {\sqrt{2}}\Gamma\cdot b_3^*|0>_{\alpha}  \\
\end{array}
\right )
$
&
$
\begin{array}{c}
1/2\\
-1/2\\
\end{array}
$
&
$
\begin{array}{c}
1/3\\
1/3\\
\end{array}
$
&
$
\begin{array}{c}
2/3\\
-1/3\\
\end{array}
$&$
\left (
\begin{array}{c}
\psi_1,
~\psi_1^*\\
~~ \psi_2,
~~\psi_2^* \\
~~ \psi_3, 
~~ \psi_3^* \\
\end{array}
\right )$\\
\hline
\end{tabular}
\end{table}
\end{center}
The next problem is to supersymmetrise the duality invariance Maxwell action,
so that the manifest duality invariance is preserved. A Majorana field 
shall be represented by a pair of complex two component spinors 
$\psi^{(\alpha)}(1\leq\alpha\leq 2)$ satisfying the condition
\begin{equation}\psi^{(\alpha)*}=\sigma_2{\cal L}_{\alpha\beta}\psi^{(\alpha)},
\end{equation}
with $\sigma_i$ is the Pauli matrices.
Along with the action for the fermions,
\begin{equation}
S_F=\int d^4x \left [i\psi^{(\alpha)\dag}\partial_o\psi^{(\alpha)}-
\psi^{(\alpha)\dag}{\cal L}_{\alpha\beta}\sigma_k\partial_k\psi^{(\beta)}
\right],\label{ef}
\end{equation}
the full action is
\begin{equation}
S_C=\int d^4x \left [i\psi^{(\alpha)\dag}\partial_o\psi^{(\alpha)}-
\psi^{(\alpha)\dag}{\cal L}_{\alpha\beta}\sigma_k\partial_k\psi^{(\beta)}
-\frac{1}{2}(B^{(\alpha)i}{\cal L}_{\alpha\beta}
E_i^{(\beta)}+B^{(\alpha)i}B^{(\alpha)i} \right ].
\end{equation}

The above action is invariant under the following supersymmetric
transformation,
\begin{equation}
\delta\psi^{(\alpha)}=\frac{1}{2}({\cal L}_{\alpha\beta}\sigma_k
B^{(\beta)k}\epsilon-\sigma_k B^{(\alpha)k}\sigma_2\epsilon^*),
\end{equation}
and 
\begin{equation}
\delta A_i^{(\alpha)}=i\psi^{(\alpha)\dag}\sigma_i\epsilon-
i\psi^{(\beta)\dag}{\cal L}_{\alpha\beta}\sigma_i\sigma_2\epsilon^*,
\end{equation}
where $\epsilon$ is the arbitrary two  component  complex spinor.
One can write the action S, where fermions can couple to gravity thereby 
achieving general coordinate invariance and Lorentz invariance,
\begin{equation}
S=\int d^4x\sqrt{-g}\left (ie_o^\mu\psi^{(\alpha)\dag}D_\mu\psi^{(\alpha)}-
e_k^\mu\psi^{(\alpha)\dag}{\cal L}_{\alpha\beta}\sigma_k D_\mu\psi^{(\beta)}\right ),
\end{equation}
where $ D_\mu $ is the covariant derivative involving the spin connection. 
We can rewrite the total fermionic part of the action as
\begin{equation}
S_F=\int d^4x\sqrt{-g}\left ((ie_o^\mu\psi_a^{(\alpha)\dag}D_\mu\psi_a^{(\alpha)}-
e_k^\mu\psi_a^{(\alpha)\dag}{\cal L}_{\alpha\beta}\sigma_k 
D_\mu\psi_a^{(\beta)}\right ),
\end{equation}
where `a' run from 1 to 12 as shown in Table-\ref{tab:table1}.

Let U's be the Wilson loop integrals to construct the ordinary derivative
from covariant derivative. For color, this is
\begin{equation}
U_C(x)= exp\left( ig\int_0^x\sum_{l=1}\lambda^l V_i^l dy_i\right).
\end{equation}
The isospin phase functions can be given, if we define
\begin{equation}
Y(x)=g'\int_0^x B_i dy_i,
\end{equation}
by the following that
\begin{eqnarray}
U_Q(x)&=&exp\left(\frac{ig}{2}\int_0^x{\tau}\cdot{\bf W}_i dy_i -\frac{i}{6}Y(x)\right),\\
U(x)&=& exp\left( \frac{ig}{2}\int_0^x{\tau}\cdot{\bf W}_i dy_i -\frac{i}{2}Y(x)\right),\\
U_1(x)&=& exp\left(-\frac{2i}{3}Y(x)\right),\\
U_2(x)&=& exp\left(\frac{i}{3}Y(x)\right),
\end{eqnarray}
and
\begin{equation}
U_R(x)=exp\left( iY(x) \right ).
\end{equation}
The fermionic and vector fields can then be put in a simpler form.
If, let
\begin{displaymath}
\psi^a =\left \{ 
\begin{array}{ll}
U_Q~U_C \left (
\begin{array}{c}
u^a_L\\
d^{a+3}_L\\
\end{array}
\right ), &{a=1,2,3};\\
U_1~ U_C~ u^a_R,&{a=4,5,6};\\
U_2~ U_C ~d^a_R, & {a=7,8,9};\\
U_r~e^a_R,&{a=10};\\
U\left(
\begin{array}{c}
\nu_c\\
e^-\\
\end{array}
\right ),&{a=11,12}\\
\end{array}\right.
\end{displaymath}

one can write, with the above $\psi^a$, the action $S_{CF}$ for supergravity 
and Yang Mills fields as
\begin{equation}
S_{CF}\sim\int d^4x\sqrt{-g}\left (ie_o^\mu\psi_a^{(\alpha)\dag}(\mathcal{F}_i\cdot 
\mathcal{F}_i)^a D_\mu\psi_a^{(\alpha)}-
e_k^\mu\psi_a^{(\alpha)\dag}(\mathcal{F}_i\cdot \mathcal{F}_i)^a
{\mathcal L}_{\alpha\beta}\sigma_k D_\mu\psi_a^{(\beta)}\right ).
\end{equation}

\section{Action Integral from Supergravity}\label{action}

Supergravity is a generalisation of supersymmetry. We shall use the formalism of
Cremmer et al ~\cite{Cremer78}. The action is local and we get~\cite{Mishra92}
\begin{equation}
{\cal A} =\int d^4x ~d^4\theta~e~\phi(\tilde{S}e^{2\tilde{g}V}S) +\int d^4x
~d^2\theta\left[ \Re (1/R)g(S) +{\Re} (1/R)f_{ab}(S)W^{a\alpha}W^b_\alpha
\right] .\label{eqA}
\end{equation}
$\theta$ is Grassmann variable, e is the superspace vielbein determinant needed for
getting the invariant volume in superspace. R is the chiral scalar curvature
superfield derived from the curvature 2-form in superspace. $f_{ab}(S)$ is the 
function of chiral multiplets so that $f_{ab}(S)W^{a\alpha}W^b_\alpha$ remains 
gauge invariant. a and b are the gauge indices for the adjoint representation. 
We have here three independent functions $\phi,~f_{ab}$ and g(S). The Lagrangian
so constructed is invariant under supergravity transformation. But the usual
Lagrangian for supergravity is not equation (\ref{eqA}). Subsequently we will see 
that the lagrangian of theory of relativity will emerge from it automatically because
of local supersymmetry. This is a newer feature of supergravity. The meaning of
chiral multiplet 12 in number are denoted by ($A^{(a)}_\mu, \bar{\chi}^{(a)}_\mu$),
the gauge fermions will be called $\psi^a$ and the field strength by 
$W^{(a)}_{\mu\nu}$.

The functions $\phi(\tilde{S},S)$ and g(S) can be combine to a single function
$ \phi'$,

\begin{equation}
\phi'(\tilde{S},S)=-\frac{1}{3}
\frac{\phi(\tilde{S},S)}{( \frac{g(\tilde{S})g(S)}{4})^{1/3}},
\end{equation}
and we shall consider a single function $G(A^{(a)}_\mu)$ given by
\begin{equation}
\phi'(A^{(a)}_\mu)= e^{\frac{1}{3}G(A^{(a)}_\mu)},
\end{equation}
then the Lagrangian will be given by the two functions  $G(A^{(a)}_\mu)$ and 
$f_{ab}$. When coupled to supergravity, the kinetic function and the 
superpotential loose their independent meaning. They enter only through the 
function $G$ and emphasising the fact that the scalar field space in supergravity
is very much like Kahler Manifold, $G$ denotes the Kahler like potential and Kahler like 
metric is $G^\mu_{~\nu}$. We thus have now the action depending on two functions 
$G$ anf $f_{\alpha\beta}$,
whereas the globally supersymmetric action has three functions $f,\phi$ and $g$.

With e as the determinant of the vielbein,the bosonic part of
 the Lagrangian is given as ~\cite{Mishra92}
\begin{eqnarray}
e^{-1}{\cal L}_B&=&exp(-G)\left ((3+G^a_{\mu}
((G^{ab})^{-1})^{\mu}_{~\nu} G^{b\nu})\right )\nonumber\\
&-&\frac{1}{2} g^2{\Re}(f^{-1})_{ab}(G^{c\mu}T^{a\nu}_\mu A^{(c)}_{\nu})
(G^{d\lambda}T^{b\sigma}_\lambda A^{(d)}_{\sigma}) 
-\frac{1}{2}R+G^{ab\mu}_{~~~\nu}~D_\lambda A^{(a)}_{\mu}~D^\lambda 
{A}^{(b)\nu}\nonumber\\
&-&\frac{1}{4}{\Re}f_{ab}W^{(a)}_{\mu\nu}W^{(b)\mu\nu}
+i\frac{1}{4}{\Im}f_{ab}W^{(a)}_{\mu\nu}\tilde{W}^{(b)\mu\nu},\label{eq:eL}
\end{eqnarray}
with
\begin{equation}
G^{a\mu}=\frac{\partial G~~~~}{\partial A^{(a)}_{\mu}},~~~~~~~
G^a_{\mu}=\frac{\partial G~~~~~}{\partial A^{(a)\mu}}~~~~~~\textnormal{and}~~~~~
G^{ab\mu}_{~~~~\nu}=\frac{\partial^2 G~~~~~~~~~~~}
{\partial A^{(a)}_{\mu}\partial A^{(b)\nu}}.
\end{equation}
$G^{-1}$ is the inverse of the matrix obtained from the second order
differentiation with respect to fields and their complex conjugates.
The first term in equation (\ref{eq:eL}) contains the potential due to 
matter sector and the second term due to gauge sector. We shall introduce 
the complex scalar field $\lambda$ which will form a part of G with
$\lambda =\lambda_1 +i\lambda_2$ as before. The $F$'s have also the same 
properties as before so that
\begin{equation}
\tilde{F}^{(a)\mu\nu}=\frac{1}{2}(\sqrt{-g})^{-1}\epsilon^{\mu\nu\rho\sigma}
F^{(a)}_{\rho\sigma},
\end{equation}
and 
\begin{equation}
\exp(-G)(3+G^a_{\mu}((G^{ab})^{-1})^{\mu}_{~\nu}(G^b)^{\nu}
\end{equation}
becomes proportional to
\begin{displaymath}
 g^{\mu\nu}\left[\frac{1}{2{\lambda_2}^2}\partial_\mu\lambda
\partial_\nu\bar{\lambda} +\frac{1}{8}Tr(\partial_\mu ML~\partial_\nu ML)\right ] .
\end{displaymath}
The term 
\begin{equation}
G^{ab\mu}_{~~~~\nu}~D_\lambda A^{(a)}_{\mu}~D^\lambda {A}^{(b)\nu}
-\frac{1}{4}{\Re}f_{ab}F^{(a)}_{\mu\nu}F^{(b)\mu\nu}
-i\frac{1}{4}{\Im}f_{ab}F^{(a)}_{\mu\nu}\tilde{F}^{(b)\mu\nu},
\end{equation}
becomes
\begin{equation}
\longrightarrow ~~-\frac{1}{2}\lambda_2F_{\mu\nu}^{(a)}(LML)_{ab}F^{(b)\mu\nu}+
\frac{1}{2}\lambda_2F_{\mu\nu}^{(a)}L_{ab}F^{(b)\mu\nu}.
\end{equation}
So, we have, on the whole, the same expression as equation (\ref{eq:77}). Only 
the last two terms left to be considered for bosons.

Classical gravity term will be $\frac{1}{2}R$ and the gauge field Lagrangian will 
be $-\frac{1}{4}W_{\mu\nu}\cdot W^{\mu\nu}$.

Now we write the ${\cal L}_{F,kin}$ as
\begin{eqnarray}
{\cal L}_{F,kin}&=&\frac{1}{2}\Re f_{ab}\left(-\frac{1}{2}\psi^a\gamma^\mu D_\mu\psi^b+
 \frac{1}{2}\bar{\psi}^a\gamma^\mu
\sigma^{\alpha\eta}\varphi_\mu W^{(b)}_{\alpha\eta} -\frac{1}{2}
\bar{\psi}^a\gamma_\mu\psi^b G^{c\alpha}D^\mu A^{(c)}_\alpha\right ) \\
&-&\frac{i}{8}\Im f_{ab}e^{-1}D_\mu\left( e\bar{\psi}^a\gamma_5\gamma^\mu\psi^b\right)-
\frac{1}{2}f^{c\eta}_{ab}\bar{\chi}^c_{\eta L}\sigma^{\mu\nu}W^{(a)}_{\mu\nu}\psi^b_L
-\frac{1}{4}e^{-1}\epsilon^{\mu\nu\lambda\sigma}\bar{\varphi}_\mu\gamma^5\gamma_\nu
D_\lambda\varphi_\sigma \\
&+&\frac{1}{8}e^{-1}\epsilon^{\mu\nu\lambda\sigma}\bar{\varphi}_\mu\gamma_\nu
\varphi_\lambda G^{a\alpha}D_\sigma A^{(a)}_\alpha 
-G^{ab\mu}_{~~~~\nu}\bar{\varphi}_{L\beta}\gamma^\alpha D_\alpha 
A_\mu^{(a)}\gamma^\beta \chi^{b\nu}_L\\
&-&\bar{\chi}^{a\mu}_{L}\gamma^\alpha D_\alpha A^{(b)}_{\nu}\chi^{c\kappa}_R
\left (G^{abc\nu}_{~~~\mu\kappa}
+\frac{1}{2}G^{ab\nu}_{~~\mu} G^{c}_{~\kappa}\right)
+G^{ab}_{\mu\nu}\bar{\chi}^{a\mu}_{L}\gamma^\alpha D_\alpha\chi^{\nu b}_R
+h.c.\label{eq100}
\end{eqnarray}
where
\begin{equation}
f_{\alpha\beta}^{a\mu}=\frac{\partial f_{\alpha\beta}}{\partial A_\mu^{(a)}}
\end{equation}
Here $\varphi$ denotes the gravitino and $\chi$ is the gaugino. We proceed to simplify
the above equation (\ref{eq100}). Taking  the gravitational counterpart as zero, 
we get the value
\begin{equation}
-\frac{1}{4}{\Re}\bar{\varphi}_\lambda\gamma^\mu D_\mu\varphi_\lambda,
\end{equation}
plus the  term which is not a physical contribution.
Next, the two terms of the gravitino kinetic energy are 
\begin{equation}
-\frac{1}{4}e^{\mu\nu\lambda\sigma}\bar{\varphi}_\mu\gamma_5\gamma_\nu D_\lambda\varphi_\sigma 
+\frac{1}{8}e^{-1}\epsilon^{\mu\nu\rho\sigma}
\bar{\varphi}_\mu\gamma_\nu\varphi_\rho G^{a\alpha}D_\sigma A^{(a)}_{\alpha}.
\end{equation}
There  still remains another term, which is the first term and is needed for the 
supergravity action to be invariant under duality transformation. This goes like
\begin{equation}
-\frac{1}{4} g'\bar\psi^a\gamma^\nu D_\nu\psi^a.
\end{equation}

Essentially, the above are the most important terms and we can proceed to
write the part of the fermion action $e^{-1}{\cal L}_F$ that does not contain
covariant derivative
\begin{eqnarray}
e^{-1}{\cal L}_F&=&e^{-G/2}\varphi_{\mu R}\sigma^{\mu\nu}\varphi_{\nu R}
+\frac{1}{4}e^{-G/2}G^{l\mu}(G^{-1})^{k\nu}_{~~l\mu}f^{*}_{abk\nu}\bar{\psi}^a\psi^b 
+e^{-G/2}\left (G^{ab}_{~~\mu\nu}-G^{a}_{~\mu}~G^b_{~\nu}
-G^{l}_\mu~(G^{-1})^{k\alpha}_{~l\eta}~G^{ab\eta}_{~k\alpha\nu}\right )
\bar{\chi}^{a\mu}_{L}~\chi^{b\nu}_{L}\nonumber\\
&-&\frac{1}{2}i\bar{g}G^{a}_\nu T^{bc}_aA^{{(b)}\nu}\bar{\varphi}_{\mu L}
\gamma^\mu\psi_R^c
-e^{-G/2}G^{a\nu}\bar{\varphi}_{\mu R}\gamma^\mu\chi_{La\nu}\nonumber\\
&+&\frac{1}{2}i(\Re f)^{-1}_{ab}f^{acd}_\mu\bar{g}
G^{l\nu}T^{be}_l A^{(e)}_{\nu }\bar{\chi}^\mu_{L c}\psi_{Ld}
+2i\bar{g}G^{a\mu}_{~~b\nu}T^{cd}_aA^{(d)}_\mu\bar{\psi}_{c R}\chi^{b\nu}_R\nonumber\\
&+&\frac{1}{32}(G^{-1})^{\mu k}_{~l\nu}f^{ l}_{ab\mu}f^{*\nu}_{cd k}
\bar{\psi}^a_L\psi^b_L\bar{\psi}^c_R\psi^d_R
+\frac{3}{32}\left (\Re f_{ab}\bar{\psi}^a_L\gamma_\mu\psi^b_R \right)^2
+\frac{1}{8}\Re  f_{ab}\bar{\psi}^a \gamma^\mu
\sigma^{\rho\sigma}\varphi_\mu\bar{\varphi}_\rho\gamma_\sigma\psi^b\nonumber\\
&+&\frac{1}{2}f^c_{ab\alpha}
\left ( \bar{\chi}^{\alpha}_{Lc}\sigma^{\mu\nu}\psi^a_L
\bar{\varphi}_{\nu L}\gamma_\mu\psi^b_R 
+\frac{1}{4} \bar{\varphi} _{\mu R}\gamma^\mu\chi^\alpha_{Lc}\bar{\psi}^a_L\psi^b_L\right )
+\frac{1}{8}G^{a\mu}_{~~b\nu}\bar{\chi}^\nu_{a R} \gamma_{\delta}\chi^b_{L\mu}
\left (\epsilon^{\alpha\beta\gamma\delta}\bar{\varphi}_\alpha\gamma_\beta\varphi_\gamma 
-\bar{\varphi}^\alpha\gamma^5\gamma^\delta\varphi_\alpha \right)\nonumber\\
&+&\frac{1}{16}\bar{\chi}^\alpha_{La}\gamma^\mu\chi^{b\beta}_R\bar{\psi}^c_R
\gamma_\mu\psi^d_L 
\left (-2G^{a}_{b\alpha\beta}\Re~ f_{cd}
+\Re~ f^{-1}_{gh} f^a_{g c\alpha} f^{*}_{hbd\beta}\right )\nonumber\\
&+&\frac{1}{16}\bar{\chi}^\mu_{La}\chi^\nu_{Lb}\bar{\psi}^c_L\psi^d_L 
\left (-4G^{ab\beta}_{e\mu\nu}~(G^{-1})^{e\alpha}_{g\beta}
f^g_{c d\alpha}+4f^{ab}_{cd\mu\nu}-
+\Re~ f^{-1}_{gh} f^a_{gc\mu} f^{*b}_{hd\nu}\right )\nonumber\\
&-&\frac{1}{16}\bar{\chi}^\alpha_{La}\sigma_{\kappa\theta}\chi^\beta_{Lb}
\bar{\psi}^c_L\sigma^{\kappa\theta}
\psi^d_L \Re~ f^{-1}_{gh} f^a_{g c\alpha} f^b_{h d\beta}
+\left( -\frac{1}{2}G_{cd\mu\nu}^{ab\alpha\beta}
+\frac{1}{2}G^{ab\delta}_{g\mu\nu}(G^{-1})^{g\eta}_{h\delta}
G^{h\alpha\beta}_{cd\eta}
-\frac{1}{4}G^{a\alpha}_{c\mu}G^{b\beta}_{d\nu}\right)
\bar{\chi}^\mu_{La}\chi^\nu_{Lb}\bar{\chi}^c_{R\alpha}\chi^d_{R\beta} \nonumber\\
&+& h.c.
\end{eqnarray} 
This is the most general form of the Supergravity Lagrangian. We have taken the help of 
Refs.-\cite{Mishra92} and \cite{Niles} who have written this Lagrangian for a simpler
processes. At a first sight, this Lagrangian
appears nonrenormalisable. However, we shall show that there exists a possibility that it will
lead to normalisable theory. 

\section{Renormalisability of Graviton, Gravitino and future outlook}\label{renorm}

We note that the symmetric, traceless tensor, which is non zero, 
physical and having zero mass, is
\begin{equation}
h_{\mu\nu}(p)=\sum_{i,j}~c_{ij}
\left ( b_{-\frac{1}{2}\mu}^{i\dag}~b_{-\frac{1}{2}\nu}^{j\dag}
+b_{-\frac{1}{2}\mu}^{i\dag}~b_{-\frac{1}{2}\nu}^{j\dag}
- \frac{1}{2} \eta_{\mu\nu}b_{-\frac{1}{2}\alpha}^{i\dag}~b_{-\frac{1}{2}}^{\alpha j\dag}
\right )|0,p\rangle.\label{}
\end{equation}
This can be taken as the graviton with the commutator;
\begin{equation}
\left [h^{\mu\nu}(p),~h^{\lambda\sigma}(p')\right ]=f^{\mu\nu\lambda\sigma} |c|^2
\delta^{(4)}(p-p'),
\end{equation}
where 
\begin{equation}
f^{\mu\nu\lambda\sigma} =g^{\mu\lambda}g^{\nu\sigma}+g^{\mu\sigma}g^{\nu\lambda}
-g^{\mu\nu}g^{\lambda\sigma}.
\end{equation}
$c_{ij}$ and $|c|^2$ include renormalisation factors.
This traceless tensor also satisfies $L_oh^{\mu\nu}(p)=-ph^{\mu\nu}=0$, and in flat 
space time,  we have~ $\Box h^{\mu\nu}(x)=0$, so that, we can take plane wave 
solution i.e. $ h^{\mu\nu}(x)= h^{\mu\nu}(p)e^{ipx}$. 
Then, we should have $h^{\mu\nu}h_{\mu\lambda}=\delta^\nu_{~\lambda} $.

The covariant derivative $D_\mu$, in terms of $\omega_\mu^{ab}$, is such that
\begin{equation}
D_\mu e^a_\lambda=0=\left (\partial_\mu+\omega_\mu^{ab}
\sigma_{bc}\right )e^c_\lambda,
\end{equation}
where $\sigma_{ab}$ is usual the antisymmetric product of two $\gamma$ matrices.

We have $g_{\mu\nu}=e_\mu^ie_\nu^j\eta_{ij}$ and $e_\mu^ie^{\mu j}=\delta^{ij}$, 
so that in the tangent space,
 \begin{equation}
h_{\mu\nu}=e_\mu^ae_\nu^bT_{ab}.
\end{equation}
After some algebra, we have

\begin{equation}
\left [ D_{\mu}, D_{\lambda} \right ] h^{\lambda}_{\nu} 
=e_{a}^{\lambda}~e_{\nu b}~R^{ac}_{\mu\lambda}T^{cb},
\end{equation}
so the Riemannian Curvature Tensor is 
\begin{equation}
R^{ab}_{\mu\lambda}=\partial_{\mu}\omega_{\lambda}^{ab}+\omega_\mu^{ac}\omega_\lambda^{cb}-
(\mu \leftrightarrow\lambda).\label{eq21}
\end{equation}
Inversion of equation (\ref{eq21}), 
\begin{equation}
\left [ D_{\mu}, D_{\lambda} \right ] h^{\lambda}_{\nu} =R_{\mu\lambda}h^{\lambda}_{\nu}
\end{equation}
is the parallel transport equation. 

In flat space time $[\partial_\mu,~\partial_\nu]$=0, this means 
$[\partial_\mu,~\partial_\nu]h^{\mu\nu}$=0. Using  the relation $\Box h^{\mu\nu}$=0, 
one can easily find that $R_{\mu\nu}=h_\nu^{\lambda\dag} R_{\mu\sigma}h^{\sigma}_{\lambda}
=h^{\lambda\dag}_\nu\left [ D_{\mu}, D_{\sigma} \right ] h^{\sigma}_{\lambda}$,
so that $[D_\mu,~D_\nu]=0$ everywhere in flat or curved space time,
\begin{equation}
R_{\mu\nu}=0.
\end{equation}
The graviton action, with $R=g^{\mu\nu}R_{\mu\nu}$, is
\begin{equation}
S_{graviton}=-\frac{1}{2\kappa}\int d^4x~e~R.
\end{equation}
This is given by the Lagrangian 
\begin{equation}
L_G=-\frac{1}{2\kappa}\sqrt{-g}~R.
\end{equation}
Here $\kappa =\sqrt{8\pi \cal{G}}$, where ${\cal G}$ is the universal gravitational constant.

Similarly, we can take, for the flat space time dimensions, the Rarita-Schwinger equation for the
objects
\begin{equation}
\epsilon^{\mu\nu\lambda\sigma}\psi_\mu\gamma^5\gamma_\lambda\partial_\nu
\psi_\sigma^{\frac{3}{2}},
\end{equation}
which is equivalent to in the "{\bf off shell momentum space}" and is given by the equation
\begin{equation}
p^{\mu}\psi_{\mu}^{\frac{3}{2}}=\gamma^{\mu}\psi_{\mu}=\gamma\cdot \psi=0.
\end{equation}
Let the ground state in Ramond-sector be
\begin{equation}
|\phi_0>=\alpha_{-1}^{\mu}|0,p> u_{\mu}~~~~\textnormal{or}~~~~~ 
|\phi_0> =  D_{-1}^{\mu}|0,p> u_{1\mu};
\end{equation}
u and $\textnormal{u}_{1\mu}$ are spinors. In either case, $L_0|\phi_0>=0$, so that this 
state is massless.
Constraints $L_o|\phi_o>$=0 and $F_1|\phi_o>$=0 imply that
\begin{equation}
p^{\mu}~u_{\mu}=0,~~~~~~~~\gamma^{\mu}~u_{\mu}=0
\end{equation}
Since $D_o^{\mu}\sim \gamma^{\mu}$, the Dirac gamma matrices and 
$\alpha_0^{\mu}=p^{\mu}$. So
$p^{\mu}u_{1\mu}$=0  and $\gamma^{\mu}u_{1\mu}$=0 as well. There is the freedom to 
choose a four component spinor $\psi^{\mu}$ such that $u_{1\mu}=\gamma^5 u_{\mu}$ 
and by this, we have
\begin{equation}
\psi^{\frac{3}{2}}_\mu=(1+\gamma_5)u_\mu,\label{eq:a}
\end{equation}
which satisfy 
\begin{equation}
\gamma^\mu\psi_\mu^{\frac{3}{2}}=p^\mu\psi_\mu^{\frac{3}{2}}=0.
\end{equation}
Furthermore, since $\gamma\cdot p\sim F_0$, we also have 
\begin{equation}
F_0\psi_\mu^{\frac{3}{2}}\sim (\gamma\cdot p)\psi_\mu^{\frac{3}{2}}=0.
\end{equation}
So, $\psi_\mu^{\frac{3}{2}} $ is the vectorial spinor of the Rarita-Schwinger equation.
This equation, or equation (\ref{eq:a}) is in accordance to calculating 
the renormalisability  condition with spin-$\frac{3}{2}$. These equations generalise to
\begin{equation}
D_\mu\psi_\mu^{\frac{3}{2}}=\left(\partial_\mu + \omega^{ab}_\mu\sigma_{ab} \right)
\psi_\mu^{\frac{3}{2}}
\end{equation}
and the Rarita-Schwinger Lagrangian becomes
\begin{equation}
L_{RS}=\frac{1}{2}\psi_\alpha\gamma_\nu\gamma^5 D_\sigma\psi_\mu^{\frac{3}{2}}
\epsilon^{\alpha\nu\sigma\mu}.
\end{equation}
The sum of the two terms $L_{RS}$ and $L_G$ is invariant under local supersymmetric 
transformation
\begin{eqnarray}
\delta^m_\mu&=&\frac{1}{2}~\kappa~ \bar{\xi}\gamma^m\psi_\mu,\nonumber\\
\delta\psi_\mu&=&\frac{1}{\kappa}\left ( \partial_\mu+\omega^{mn}_\mu\sigma_{mn} 
\right ) \xi=\frac{1}{\kappa}D_\mu \xi\nonumber\\
\delta\omega_\mu^{mn}&=&0.
\end{eqnarray}
Thus , the Lagrangian appears to be renormalizable. This will be taken up 
further in a future work.

\section{conclusion}\label{concl}
We also note that we have arrived at the group 
$Z_3\otimes SU_c(3)\otimes SU_L(2)\otimes U_Y(1)$
to explain the experimental result. It is of the interest to note that 
the number of generations $n_G$ will be 
given by the Euler number $\chi$ which is 6 in our consideration of 4-dimensional
theory, so that~~$n_G=\frac{\chi}{2}$~=~$n_+-n_-$~=~3, where $n_-$ and $n_+$ are 
the number of negative and positive chiralities respectively. The zero mass modes 
of Dirac objects in standard model are grouped into three families(generations). 
They are
\begin{eqnarray}
\left (
\begin{array}{c}
u\\
d\\
\nu_e\\ 
e\\
\end{array}
\right ),
\left (
\begin{array}{c}
c\\
s\\ 
\nu_{\mu}\\
\mu\\
\end{array}
\right )~~~\textnormal{and}~~~~
\left (
\begin{array}{c}
t\\
b\\
\nu_{\tau}\\
\tau\\
\end{array}
\right ).
\end{eqnarray}
The left handed ones are doublets and right handed
ones are singlets. The quarks are colored. There are
several fermionic zero modes, 24 are with +ve helicity and 21
with negative helicity. Thus $n_+-n_-$=3 as per topological
findings. So there have to be only three fermions (neutrinos
with unpaired helicity) in the Standard Model. 

Thus, we have explored 
the four dimensional supergravity as far as practicable. We have started
and also ended with four dimensions and never taken any more dimensions, specifially, 
nothing has been done to involve ten dimensions, as four dimensions appear to be good 
enough for all purposes. We have examined the difficulties of the four dimensional
methods. But we find that it is more efficient than ten dimensions or heteretic string
theory. Our approach to quantise gravity, using physical states of a 
superstring, has been quite successful\cite{Deo09}. It has been noted that spin-2 
graviton and spin zero dilaton are obtained from the metric field strength $g_{\mu\nu}$,
where as spin zero part can be eliminated in superstring theory. With a four dimensional 
theory, we have considered the question of renormalisability of gravitational field 
and spin-$\frac{3}{2}$ fermions afresh and find that it is quite possible for a total 4-d 
theory which includes all interactions.

\end{document}